# Micron-sized silica-PNIPAM core-shell microgels with tunable shell-to-core ratio

Keumkyung Kuk [a], Lukas Gregel [a], Vahan Abgarjan [a], Caspar Croonenbrock [a], Sebastian Hänsch [b], and Matthias Karg [a]*

[a]Institut für Physikalische Chemie I: Kolloide und Nanooptik, Heinrich-Heine-Universität Düsseldorf, Universitätsstr. 1, 40225 Düsseldorf, Germany

[b]Center for Advanced Imaging, Heinrich Heine University Düsseldorf, Universitätsstraße 1, 40225 Düsseldorf, Germany.

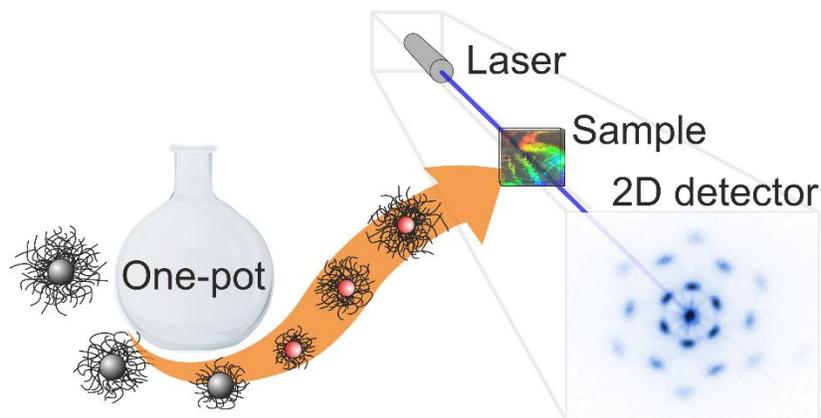

**Abstract:** Micron-sized hard core-soft shell hybrid microgels are promising model systems for studies of soft matter as they enable in-situ optical investigations and their structures/morphologies can be engineered with a great variety. Yet, protocols that yield micron-sized core-shell microgels with a tailorable shell-to-core size ratio are rarely available. In this work, we report on the one-pot synthesis protocol for micron-sized silica-poly(*N*-isopropylacrylamide) core-shell microgels that has excellent control over the shell-to-core ratio. Small-angle light scattering and microscopy of 2- and 3-dimensional assemblies of the synthesized microgels confirm that the produced microgels are monodisperse and suitable for optical investigation even at high packing fractions.

**1. INTRODUCTION**

Microgels are colloid-like, deformable and soft objects that have interior structures resembling gel-like characteristics swollen by a solvent in which they are dispersed [1, 2]. They belong to a unique class of materials because they exhibit both solid- and liquid-like behavior and have the capability to respond to external stimuli, e.g., temperature [3], pH, and ionic strength [4]. These properties can be tuned by engineering the morphologies [5], composition [6], porosity, and elasticity of the microgels [7]. Due to their tailorable stimuli sensitivity, high colloidal stability, and a broad range of possibilities for various functionalization, the past decades have seen a steadily increasing interest in microgels in applications such as biomedicine, photonic and process technology as well as in fundamental research across disciplines [1]. Among others, inorganic core-polymeric shell microgels, also known as hard core-soft shell microgels, have received significant attention, for example, because of their hybrid properties [8, 9], fine-tunable interparticle distance [10] and their potential to be assembled into surprisingly complex microstructures despite their isotropic shape [11, 12].

In general, there are two approaches to prepare such core-shell (CS) microgels: "grafting from" and "grafting to" approaches [13, 14]. In the "grafting from" approach, the polymer chains grow from the core surface, allowing precise control over the shell thickness [12] with a certain limit. On the other hand, in the "grafting to" approach, pre-formed polymer chains/gels are anchored to/adsorbed onto the core surface. A good example is the free radical precipitation polymerization, which is by far the most widely used synthesis technique that can offer a broader size range as well as various post-modification of polymeric shell morphologies and composition via semi-batch and/or seeded precipitation polymerization methods [15-17]. The post-modification can also be used for the overgrowth of the shell, increasing the overall dimension of the CS microgels, hence the shell-to-core size ratio ($\delta$). CS microgels with dimensions close to the micron regime were synthesized via multiple-step addition of monomer in the past [18, 19]. The micron-sized CS microgels could serve as a very convenient model system because their larger sizes, and thus slower diffusion, can enable in-situ optical investigations using, e.g., optical tweezers, simple light microscopy as well as small-angle light scattering (SALS), which are extensively customizable and cost/time efficient in-house methods. However, one-step synthesis for micron-sized CS microgels with tailorable $\delta$ has not been reported yet.

In this study, we present a facile and robust one-pot synthesis protocol to prepare micron-sized, monodisperse CS microgels with controllable $\delta$ via a surfactant-free seeded precipitation polymerization. We chose silica as the core material because it is generally biocompatible and can be synthesized with great control over size with low polydispersities, not to mention its facile control over its pore sizes and surface properties via simple silanization chemistry [20, 21]. We chose poly(*N*-isopropylacrylamide) (PNIPAM) as the shell material not only because it is one of the most commonly used and well-studied polymers but also because of its thermoresponsive nature around ambient temperature, which could give rise to numerous applications. We have covered core sizes ranging from 245 to 455 nm in diameter with overall hydrodynamic diameter ($D_h$) of the CS microgels up to approximately 1.2 μm. The swelling capacity of the different samples was studied by temperature-dependent dynamic light scattering (DLS). All prepared colloids show thermoresponsive properties in water due to the lower critical solution temperature (LCST) behavior of PNIPAM. As a proof of concept, the synthesized microgels are assembled into 2D and 3D microstructures. These superstructures were successfully studied by SALS, light microscopy as well as confocal microscopy.

## 2. RESULTS & DISCUSSION

**2.1. Synthesis and characterization.** We used the seeded precipitation polymerization to synthesize CS microgels with differently sized silica cores and thermoresponsive PNIPAM shells of different thicknesses. For such synthesis, silica cores are commonly surface-functionalized with methacrylates to establish covalent bonds to the PNIPAM shell. Especially in the high total solids content (TSC, here defined as the mass of all the suspended and dissolved solids in the sample divided by the total volume of continuous phase—water) regime, this hydrophobic surface modification can hamper colloidal stability during the synthesis, which can lead to mixed species with double/triple cores, high polydispersity or macroscopic aggregates. Previously, Karg et al. reported significantly lower encapsulation rates with increasing size of the silica cores [22]. In that study, silica cores with sizes ranging from 68 nm to 170 nm in diameter were used. In more recent studies, silica cores with a size of 351 nm were successfully encapsulated in PNIPAM shells leading to micron-sized CS microgels ($D_h \approx 1$ μm) via seeded precipitation polymerization with multiple-step monomer addition for overgrowing of the shell in the presence of surfactant [19, 23].

Others have also reached microgel dimensions on the order of 1 µm via continuous feeding of the monomers in the absence of surfactant [18]. However, precise control over the δ targeting overall dimensions reaching the micrometer regime still seems to be challenging, in particular for one-pot reactions, also known as single batch polymerizations. In this study, we aim to tackle this challenge and propose a simple but robust synthesis route for micron-sized CS microgels that are sufficiently large to be suitable for investigations using light—either in optical microscopy or diffraction. For the preparation of micron-sized CS microgels, we have found that the shell growth is significantly more effective in the lower TSC regime in the absence of surfactant under efficient stirring. Figure 1 shows representative transmission electron microscopy (TEM) images for small (245 nm), medium (388 nm), and large (455 nm) silica cores and the corresponding CS microgels with different shell thicknesses synthesized via surfactant-free precipitation polymerization at low TSC (the proposed synthesis protocol, more details can be found in the Section 4). All CS microgels are labeled using $C_xS_y$, where x represents the silica core diameter as determined from TEM and y corresponds to the shell-to-core size ratio, δ. δ is defined as $D_h$ measured by dynamic light scattering at 20 °C (swollen state) divided by the diameter of the silica core measured by TEM, for example, 245 nm core with δ of 2.9 ($C_{245}S_{2.9}$) and 455 nm core with δ of 2.3 ($C_{455}S_{2.3}$). All δ values were calculated with $D_h$ as acquired and without error propagation, see Supplementary information (SI) for more details. The differently sized silica cores in (Figure 1A–C) possess spherical shape and low dispersity in size. In images (Figure 1D–I), the lower contrast area on the edge of the silica cores evidently shows that the PNIPAM encapsulation was successful. In particular, for the microgels with the thickest shells (Figure 1H,I), the shells are clearly visible. Here, each silica core is surrounded by a homogeneous PNIPAM shell. For the microgels with thinner shells, the boundary between the higher electron density—rigid cores, and the low electron density polymeric shell is less noticeable but clearly visible at higher magnification. The TEM images with higher magnifications can be found in SI, Figure S1. We want to note that the samples are imaged in the dried state and under high vacuum conditions in the TEM. Consequently, the shells are imaged in a collapsed state with a much smaller dimension than in bulk dispersion, when the shells are swollen with water.

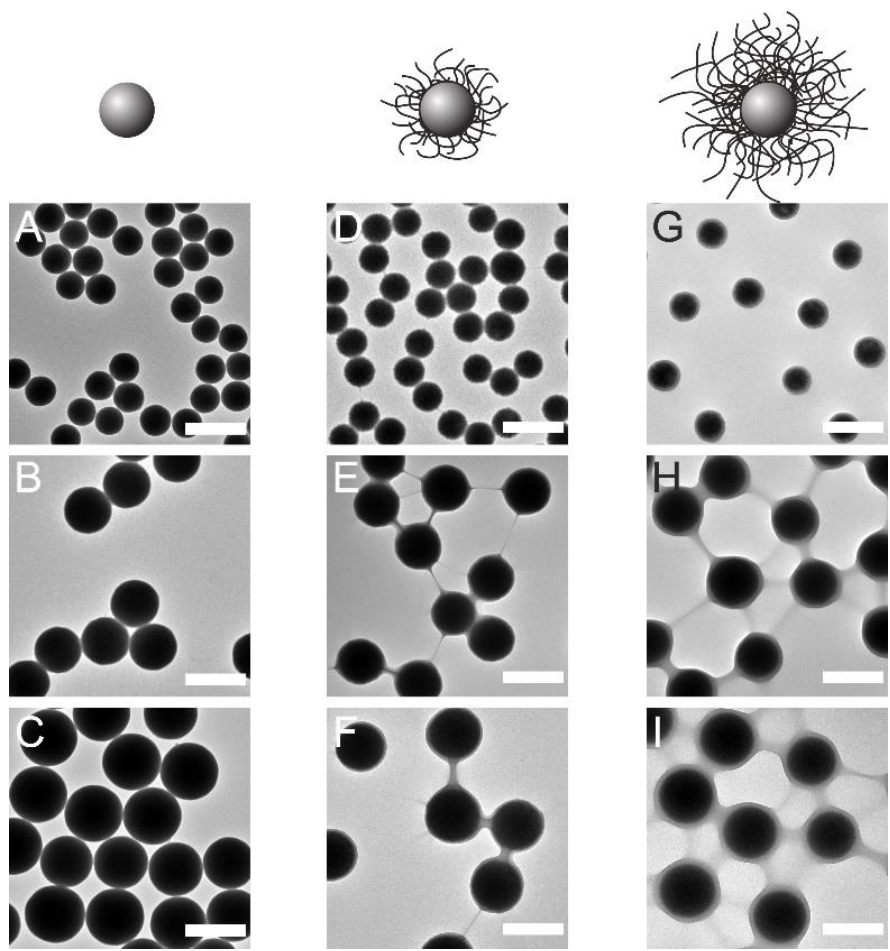

**Figure 1.** TEM images of CS microgels with variously sized cores: $C_{245}$ (**A**), $C_{388}$ (**B**), and $C_{455}$ (**C**), with thin shells: $C_{245}S_{1.7}$ (**D**), $C_{388}S_{2.1}$ (**E**), $C_{455}S_{2.1}$ (**F**) and with thicker shells: $C_{245}S_{2.9}$ (**G**), $C_{388}S_{2.6}$ (**H**), $C_{455}S_{2.3}$ (**I**). The scale bars correspond to 500 nm.

Figure 2A shows a 3D plot of shell growth on $C_{245}$ in terms of δ as a function of NIPAM concentration per number of core and TSC. The blue spheres represent data for CS microgels synthesized without surfactant in the low TSC regime, using an overhead stirrer (the proposed synthesis protocol), while the green tetrahedrons correspond to results from synthesis with surfactant (sodium dodecyl sulfate-SDS, 2 mM) in the high TSC regime, using a magnetic bar for stirring (following the protocol from [24], labeled as $C_xS_y$-C for Core-Shell-Conventional). More details on the effects of the individual synthesis parameters such as SDS concentration, stirring method, and temperatures on the overall size of the CS microgels can be found in SI. Figure 2B shows the shell growth in terms of δ as a function of the NIPAM concentration per number of silica cores for the three different core sizes (see Figure 1) performed via the proposed synthesis protocol at low TSC in the absence of SDS under more efficient stirring. The results show that the synthesis protocol can also be transferred to significantly larger cores. The shell growth of all CS microgels can be described by an exponential growth as a function of monomer concentration per number of cores (fitted with the Gompertz fit, solid lines). This enables us to predict the total microgel sizes for any given feed concentration, at least in the studied range. At the same time, desired values of δ can be specifically targeted. Figure 2C shows the swelling curves of the four selected CS microgels recorded by temperature-dependent DLS: CS microgels with the small core with thin shell $C_{245}S_{1.7}$ (filled blue) and thick shell $C_{245}S_{2.9}$ (empty blue) and the large core with thin shell $C_{455}S_{2.1}$ (filled red) and thick shell $C_{455}S_{2.3}$ (empty red). In all cases, we find the typical volume phase transition (VPT) behavior with a continuous decrease in hydrodynamic diameter with increasing temperature in the vicinity of the volume phase transition temperature (VPTT). Here, we want to note that our larger CS microgels seemed to be under the effect of gravitational settling during the DLS measurements, which could be the reason for their higher polydispersities (or overestimation of polydispersity) compared to the smaller microgels. Although the commonly quoted upper size limit for DLS is around 10 μm, it often is only achievable by increasing the viscosity of the continuous phase or by using capillary DLS [25, 26]. In this study, however, we do not further discuss the matter and report the $D_h$ values as recorded and used for the calculation of δ. The calculated de-swelling ratios (α) of the corresponding CS microgels appear to overlap rather well as depicted in Figure 2D. The de-swelling ratios α were calculated as:

$$\alpha = \frac{V_{cs}(T) - V_c}{V_{cs}(10\ °C) - V_c} \quad (1)$$

where $V_{cs}(T)$ denotes the volume of the total CS microgel measured by DLS at temperature $T$, $V_c$ the volume of the non-swellable and non-responsive silica cores measured by TEM.

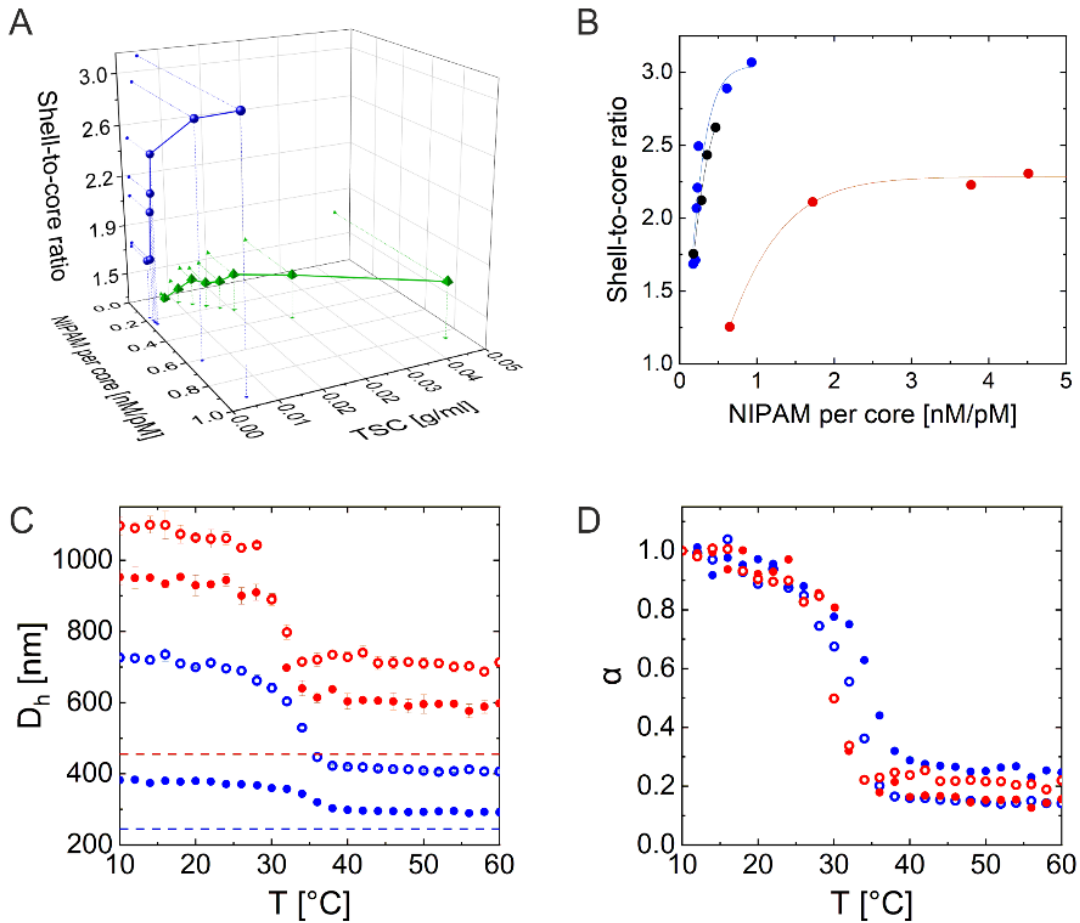

**Figure 2**. (**A**) 3D plot of shell growth on $C_{245}$ measured by DLS (depicted as shell-to-core ratio) in dependence of the NIPAM concentration per number of cores in nM/pM and TSC in g/mL. Blue: $C_{245}$ encapsulation via proposed protocol, green: according to [24]. (**B**) Shell growth on differently sized cores illustrated as the shell-to-core ratio in dependence of the NIPAM feed concentration per number of cores in nM/pM. Blue: $C_{245}$, black: $C_{388}$, red: $C_{455}$. The solid lines correspond to the Gompertz fit of the growth. (**C**) Swelling curves of $C_{245}S_{1.7}$ (filled blue), $C_{245}S_{2.9}$ (empty blue), $C_{455}S_{2.1}$ (filled red) and $C_{455}S_{2.3}$ (empty red). The dotted lines represent the diameter of the core measured by TEM. (**D**) Calculated deswelling ratio of corresponding CS microgels from (**C**).

### 2.2. Investigation of 2D assemblies using optical microscopy and SALS.

To study 2D assemblies of the CS microgels, we prepared hexagonally ordered monolayers using interface-mediated self-assembly and subsequent transfer to glass substrates. The monolayer assembly is also an effective way to judge their collective behaviors as well as the monodispersity. The samples were transferred to the substrates at surface pressures of approximately 20 mN/m, i.e., at relatively high pressures, where the CS microgels are already in shell-shell contact and squeezed against each other. More information on the sample preparation can be found in the Section 4 and in SI. We prepared monolayers from CS microgels with the smallest ($C_{245}S_{2.9}$) and largest cores ($C_{455}S_{2.3}$) investigated in this study. Figure 3B,E show optical light microscopy images of the samples at 100× magnification. For both samples, this magnification is clearly sufficient to resolve single particles. Furthermore, the hexagonal order becomes evident, which is also reflected by the six-fold symmetric fast Fourier transformations (FFTs) shown in (Figure 3A,D). While the microscopic images cover areas on the order of 0.01 mm², we can probe significantly larger areas over 1 mm² when using SALS on the same samples. Figure 3C,F show the recorded diffraction patterns. In both cases, we again see six-fold symmetries, which are in very good agreement with the FFTs computed from the real space images. Thus, optical microscopy and SALS can deliver complementary information despite the different areas probed. We want to highlight that typically microgel and CS microgel assemblies were studied using scattering techniques based on neutrons and/or X-rays (mostly SANS and SAXS) in the past years and/or by rather high-resolution microscopies such as AFM and scanning electron microscopy (SEM). Being able to use light for the structural investigation offers great possibilities for investigation of microstructures and phase transitions in real time being independent of large-scale facilities and expensive setups.

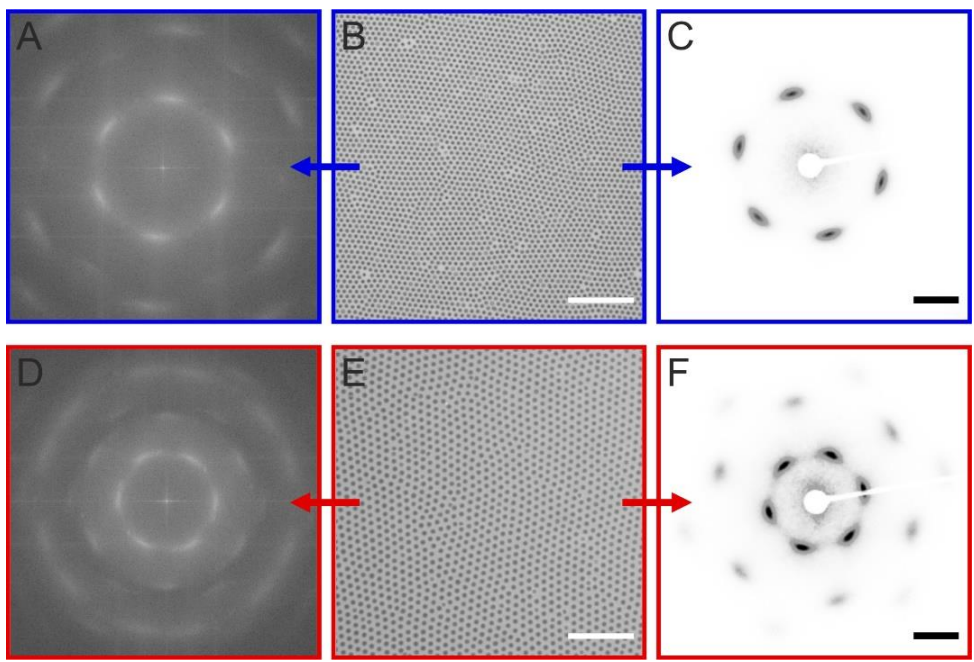

**Figure 3.** Monolayer analysis by optical light microscopy and SALS. (**A**) FFT generated from the real space microscopy image of a monolayer prepared from $C_{245}S_{2.9}$ taken at a surface pressure of 20 mN/m (**B**). (**C**) Corresponding diffraction pattern recorded by SALS. (**D,E**) same set of data as in the top row for a monolayer prepared from $C_{455}S_{2.3}$. Scale bars in (**B,E**) correspond to 10 μm. Scale bars in (**C,F**) correspond to 20 mm.

Since the nearest neighbor center-to-center distance, i.e., interparticle distance ($D_{c-c}$) in 2D assemblies of CS microgels with soft and deformable shells depends on the number of microgels per area, we can cover a broad range of distances with the same batch of microgels simply through adjusting the surface pressure, for example, in a Langmuir trough. Figure 4A shows swelling curves of the two selected CS microgels recorded by temperature-dependent DLS. From these data we can now estimate the possible range of $D_{c-c}$ and area fraction based on two assumptions: (1) The shell deformation at the interface leads to a microgel diameter that is larger by a factor of 1.76 than the bulk hydrodynamic diameter [24] and the minimum theoretically possible interparticle distance, i.e., $D_{c-c}$ in effective core-core contact, corresponds to the $D_h$ of collapsed CS microgels at 60 °C. (2) The microgels remain perfectly circular and hexagonally arranged throughout the compression, and the defects/empty spaces in the monolayer can simply be reflected by a lower area fraction as expressed in Equations (2) and (3).

$$A_P = \pi \times \left(\frac{D_{c-c}}{2}\right)^2 \tag{2}$$

$$n_P = \frac{A_f \times A_{tot}}{A_P} \tag{3}$$

$$D_{c-c}^2 = \frac{4}{\pi} \times A_f \times \frac{A_{tot}}{n_p} = \frac{4}{\pi} \times A_f \times \left(n_P/A_{tot}\right)^{-1} \tag{4}$$

Here, $A_P$ denotes the area occupied by one particle, $n_P$ is the number of particles, $A_f$ is the area fraction, and $A_{tot}$ is the total area. It is clear that $D_{c-c}^2$ scales linear with respect to the area per particle ($A_{tot}/n_P$), with a slope equal to $4/\pi$ multiplied by $A_f$. $A_f$ thus can be estimated from the real images. More details on the $A_f$ calculation can be found in SI. Based on these assumptions, we calculated the achievable range of $D_{c-c}$ as a function of the particle number per unit area ($n_P/A$) in (Figure 4B) (solid lines) at maximum $A_f$ (0.91) for the perfect hexagonal arrangement. The diagram also contains measured values of $D_{c-c}$ (symbols) from monolayers taken from the air/water interface at surface pressures of 10, 20, and 30 mN/m. Note that the measured $D_{c-c}$ lie slightly lower at the same $n_P/A$ values compared to the calculated $D_{c-c}$, likely due to occasional non-ideal packing or defects of the monolayers. The estimated area fractions are 0.80 for $C_{245}S_{2.9}$ and 0.79 for $C_{455}S_{2.3}$, respectively.

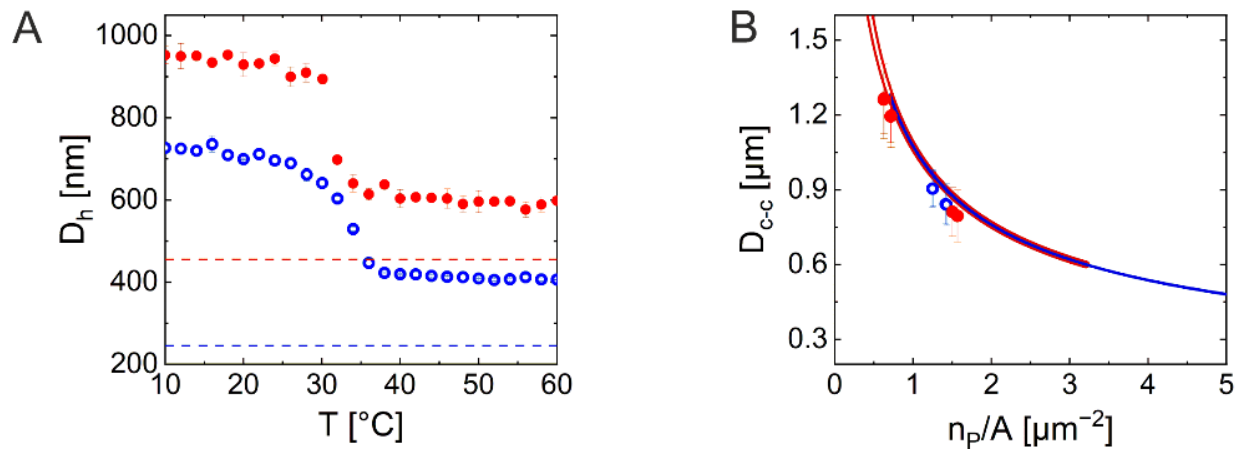

**Figure 4**. (**A**) Evolution of the hydrodynamic diameter ($D_h$) for $C_{245}S_{2.9}$ (blue) and $C_{455}S_{2.3}$ (red) as a function of temperature. The dashed, horizontal lines correspond to the respective core diameters measured by TEM. (**B**) Calculated interparticle distance ($D_{c-c}$) against the number of particles per unit area ($n_P/A$) for 2D compression accounting for an area fraction of 0.91 (solid lines, blue: $C_{245}S_{2.9}$, red: $C_{455}S_{2.3}$). Symbols with error bars: measured data at three different surface pressure (10, 20, and 30 mN/m, blue: $C_{245}S_{2.9}$, red: $C_{455}S_{2.3}$).

**2.3. Investigation of 3D assemblies using confocal microscopy and SALS.** Having shown that our CS microgels are suitable for structural investigations of monolayer samples using light, we now want to turn to 3D assemblies. For this, we chose the sample $C_{340}S_{3.0}$ (silica core dyed with rhodamine B, see SI, Tables S1 and S2 for more details) and prepared variously concentrated dispersions using *N*-methyl-2-pyrrolidone (NMP) as solvent. NMP was chosen to reduce the scattering contrast as NMP has a refractive index (1.47) higher than water (1.33) almost matching the index of PNIPAM (1.50) and silica particles (~1.45). This way we could reduce multiple scattering which was necessary for the sample investigation by SALS. The samples were sealed in glass capillary tubes; more details can be found in the Section 4. Figure 5 shows confocal laser scanning microscopy (CLSM) images obtained from $C_{340}S_{3.0}$ NMP dispersions with two different concentrations. The imaging by CLSM revealed that the sample with the lower concentration had an interparticle distance close to the $D_h$ of its building block with periodical spatial arrangements (Figure 5B), whereas the more concentrated sample showed shorter interparticle distances with amorphous structures (Figure 5E). Both images were taken in the middle of the glass tube and evidently show different spatial arrangements between the two samples. The FFT of image (Figure 5B) has a six-fold symmetry as shown in (Figure 5A), whereas the FFT of image (Figure 5E) shows a diffraction ring as depicted in (Figure 5D), reflecting the amorphous structure. Figure 5C,D show the diffraction patterns recorded from the SALS measurement. The recorded diffraction patterns are in very good agreement with the FFTs of the confocal images. Additionally, we have also acquired z-stacks from one glass wall through the sample to the other glass wall, revealing the spatial arrangement throughout the sample. See SI for the avi files.

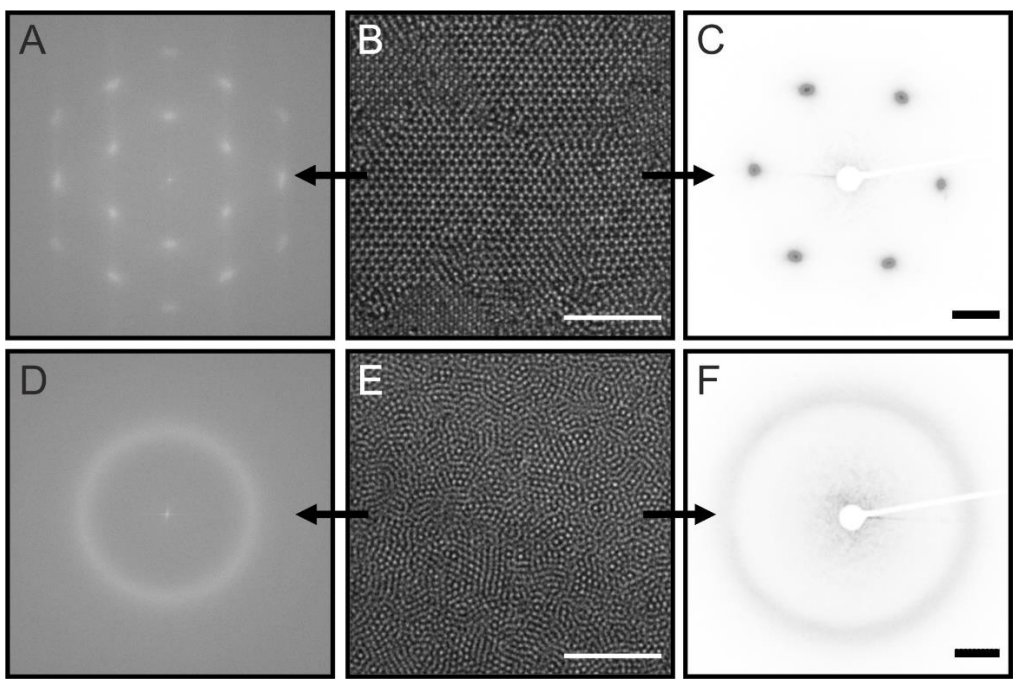

**Figure 5**. 3D colloidal microstructure analysis by CLSM and SALS (A) FFT calculated from a CLSM image of $C_{340}S_{3.0}$ dispersed in NMP (naturally sedimented from a dilute sample) taken at approximately 58 μm distance from the glass tube wall (**B**). (**C**) Corresponding diffraction pattern from SALS performed on the same sample as shown in (**B**). (**D**–**F**) the same set of data as in the top row for $C_{340}S_{3.0}$ dispersed in NMP at a higher concentration taken at approximately 50 μm distance from the glass tube wall (**E**). The scale bars correspond to 10 μm in (**B,E**) and to 20 mm in (**C,F**).

## 3. CONCLUSION

In this work, we presented an optimized synthesis protocol that uses the one-pot seeded precipitation polymerization for the preparation of micron-sized core-shell microgels. Silica particles of different sizes were used as rigid cores. The microgel shells were composed of chemically cross-linked poly-*N*-isopropylacrylamide—a thermoresponsive polymer. Due to the single batch nature, our procedure is time- and cost-efficient and offers great control over the shell-to-core ratio. As proof-of-concept experiments to demonstrate the suitability of the microgels for structural investigations using visible wavelength light, we studied substrate-supported monolayers as well as 3D samples in capillaries by confocal laser scanning microscopy and small-angle light scattering. The presented core-shell microgels are not only interesting for photonic applications but also allow convenient microstructural analyses using light rather than X-ray/neutron scattering, enabling cost-/time-efficient in-house investigations of large sample volumes. The presented systems are ideal model colloids to study, for example, interaction potentials of soft microgels with different morphologies under various conditions, their wetting/de-wetting behaviors, melting and crystallization processes as well as jamming transitions.

## 4. EXPERIMENTAL SECTION

### 4.1. MATERIALS

Ethanol (99.8%, Sigma-Aldrich, Taufkirchen, Gemany), ethanol (p.a., Heinrich-Heine-University, chemical store), tetraethyl orthosilicate (TEOS, 98%, Sigma-Aldrich), ammonium hydroxide solution ($NH_3$ (aq.), 30%, PanReac AppliChem, Darmstadt, Germany), ammonium hydroxide solution ($NH_3$ (aq.), 25%, VWR, Darmstadt, Germany), hydrogen peroxide solution ($H_2O_2$, 30 wt %, Fisher Scientific, Schwerte, Germany), *N*-methyl-2-pyrrolidone (NMP, 99.5%, Sigma-Aldrich), rhodamine B isothiocyanate (mixed isomers, Sigma-Aldrich), (3-aminopropyl)trimethoxysilane (97%, Sigma-Aldrich), 3-(trimethoxysilyl)propyl methacrylate (MPS, 98%, Sigma-Aldrich), *N*,*N*'-methylenebisacrylamide (BIS, 98%, Sigma-Aldrich), sodium dodecyl sulfate (SDS, Ph. Eur., Merck, Darmstadt, Germany), and potassium peroxodisulfate (KPS, 99%, Sigma-Aldrich) were used as received. Water was purified by a Milli-Q system (18.2

MΩ cm) and *N*-isopropylacrylamide (NIPAM, 97%, TCI) by recrystallization from cyclohexane (99.8%, Fisher Scientific).

**4.2. SYNTHESIS**

**4.2.1. Synthesis and surface modification of colloidal silica cores.** The silica cores were synthesized via the well-known Stöber procedure [27] and surface-modified with MPS. The details of the synthesis protocol have been reported elsewhere. [24] The synthesis conditions and chemicals used are listed in SI, Table S1. Purification of the synthesized silica particles was performed by repeated centrifugation for 2–8 min at 5000–7000 rcf and re-dispersion in ethanol until the supernatant cleared and the smell of ammonia vanished. The concentrated dispersion was stored in ethanol on a 3D shaker. The TSC of the dispersion in g/mL—the amount of solids remaining after storing in the oven at 80 °C for 4 h—was measured three times and averaged. The particle density was assumed to be 2.1 g/mL for the estimation of particle number concentrations [28]. The size of the silica particles was measured by transmission electron microscopy (TEM). The particles had diameters ranging from 245 to 455 nm with polydispersities on the order of 3.5–9.0%.

**4.2.2. Synthesis of silica-PNIPAM CS microgels at fixed temperature.** CS microgels were synthesized using seeded precipitation polymerizations with various feed concentrations of NIPAM as monomer and fixed ratios of the cross-linker BIS of 5 mol% (with respect to NIPAM). Eight different silica-PNIPAM CS microgel systems were synthesized following a previously published protocol using an oil bath (silicon oil) heated to 80°C. [24] Additional CS microgels were synthesized with modified protocols described in the following: The corresponding amounts of NIPAM and BIS were dissolved in water in three-neck round-bottom flasks equipped with a reflux condenser and an overhead stirrer (KPG). The mixtures were heated in an oil bath to oil temperatures of 60–80 °C and purged with argon for one hour while stirring at the speed of 250–300 rpm. Then the respective volumes of silica core stock dispersions were added and the mixtures were further purged with nitrogen to remove oxygen. After the target temperature was reached and stabilized, the polymerizations were initiated by the rapid addition of aqueous 0.01 wt% KPS solution. The polymerizations were allowed to proceed for at least three hours. The final dispersions were hot-filtered through glass wool in a funnel and purified by repeated centrifugation for 2–8 min at 5000–7000 rcf and re-dispersion in water until the supernatant cleared. The purified dispersions were either freeze-dried for 3D assembly experiments or solvent-exchanged against ethanol via repeated centrifugation and re-dispersion in ethanol for 2D assembly experiments. Further syntheses with variation in SDS and KPS concentrations were performed on smaller scales (approx. 6 mL in total volume). All samples that were synthesized and a detailed list of all synthesis conditions and amount of materials are provided in the SI, Table S2.

**4.2.3. Synthesis of silica-PNIPAM CS microgels using a temperature ramp.** A well-established temperature-ramp, surfactant-free precipitation polymerization synthesis protocol [17] was modified for the seeded polymerization with silica cores as seeds. The same setup was used as for the synthesis with the fixed temperatures described above. The reaction mixtures were equilibrated at 45 °C and purged with argon for one hour while stirring at the speed of 300 rpm. After the initiation, as soon as the appearance of turbidity was visually detected, the temperature was increased to 65 °C at the average rate of 12.6 °C/h. The polymerization was allowed to proceed for three more hours after the final temperature was reached. The microgels were purified as for the protocols at fixed temperatures.

## 4.3. METHODS

**4.3.1. Monolayer preparation.** Prior to the monolayer preparation, the glass substrates (Premium microscope slides 12-544-4, Fisher Scientific) were cut in six smaller pieces (width = 13 mm, length = 25 mm), rinsed with water and placed in a customized glass holder in a beaker for RCA-1 cleaning. The glass slides were treated in $H_2O/NH_4OH/H_2O_2$ solution with a volume ratio of 5:1:1 at 80 ± 5 °C for 20 min [29, 30]. The monolayer was created by injecting microgel dispersions (in ethanol) directly to the air/water interface in a crystallizing dish filled with water. Three different surface pressures (approximately 10, 20 and 30 mN/m) were targeted to vary the interparticle distance of the transferred monolayer by injecting certain volumes of microgel dispersion. The monolayer was then transferred on to the glass substrate by pushing the glass substrate through the monolayer at the edge of the crystallizing dish and lifting up at the center at a steep angle (70–90°) and dried. Further details can be found in the SI.

**4.3.2. Preparation of 3D assemblies.** Rectangular hollow glass tubes (5012, path length = 0.1 mm, width = 2 mm, VitroCom, Mountain Lakes, USA) were used to prepare 3D colloidal microstructures in various regimes. The concentrated dispersions were prepared from freeze-dried CS microgels re-dispersed in NMP by repeated overnight shaking (7-0045, neoLab, Heidelberg, Germany) and ultrasonication. The resulting viscous dispersion was sucked into the glass tube by an aspirator. A combination of a 10–200 μm micropipette tip and a piece of thin parafilm was used as a flexible connector between the glass tube and the aspirator. When the dispersion occupied about two thirds of the tube, the open end was sealed by using an oxyhydrogen torch. The tube containing the dispersion was then flipped upside down, cleaned and centrifuged gently. After all dispersions migrated to the sealed bottom, the other end was also sealed by the torch. The prepared samples were vertically stored until no further sedimentation was observed. The dilute samples sedimented and formed strongly iridescent colloidal crystals, whereas concentrated samples did not show any visible changes.

**4.3.3. Dynamic light scattering (DLS).** The hydrodynamic diameter, $D_h$, of the CS microgels were determined using a Zetasizer Nano S (Malvern Panalytical, Kassel, Germany). The device is equipped with a HeNe laser (4 mW, 633 nm) along with a temperature-controlled jacket. Measurements were performed at a scattering angle of 173° in the temperature range between 10 and 60 °C in steps of 2 °C. Three measurements (per temperature and sample) were performed using samples filled in semi-micro PMMA cuvettes (634-0678, VWR) with 10 min of equilibration duration at each temperature. Values of $D_h$ reported are z-average values as obtained from the measurement software. All samples had polydispersities (PDI, polydispersity index) on the order of 5–10% on average for small microgels and 10–30% for micron-sized microgels with standard deviations up to 7%.

**4.3.4. Small-angle light scattering (SALS).** Diffraction patterns were recorded by a self-built setup. A blue laser (LDM-20-405, 20 mW, 405 nm, MediaLas, Balingen, Germany) was used as a light source and the images were captured in the dark with a CCD camera (DCU223C-MVL6WA, Thorlabs, Bergkirchen, Germany) and a paper screen as a detector. Acquired images were grey scaled and inverted with ImageJ (1.53 k, National Institutes of Health, Bethesda (Maryland), USA) for better visibility. Further details can be found in the SI.

**4.3.5. Transmission electron microscopy (TEM).** TEM measurements were performed using a JEOL JEM-2100 Plus (JEOL GmbH, Freising, Germany) microscope operated in bright-field mode at 80 kV acceleration voltage. Samples of the silica cores were prepared by applying a drop of an ethanolic particle dispersion on a carbon-coated copper grid (200 mesh, Science Services, Munich, Germany) and drying at room temperature. Samples of the CS microgels were prepared by transferring the monolayer from the air/water interface onto the copper grids, as described for the monolayer preparation. All images were subsequently processed using ImageJ.

**4.3.6. Optical light microscopy.** A Nikon Eclipse LV150N equipped with a 100× objective was used to acquire images of the microgel monolayers on the glass substrates. At least two images at different positions were recorded per sample. ImageJ was used to perform fast Fourier transformations (FFT) on the acquired images as well as to find the radial distribution function from the detected particle centers (macro version: 2011-08-22 by Ajay Gopal).

**4.3.7. Confocal laser scanning microscopy (CLSM).** 3D samples in the glass tubes were imaged using a Zeiss LSM880 Airyscan microscope system (Zeiss Microscopy GmbH, Jena, Germany), equipped with a Plan-Apochromat 63×/1.4 oil immersion objective lens. The sample tubes were mounted on the stage on a glass slide-supported 3D-printed adapter fitting in the object holder of the motorized stage. The silica cores were covalently labeled with Rhodamine B so the samples were imaged using a 561 nm excitation laser and a BP 570–620 + LP 645 emission filter. The acquisition was performed in airyscan super-resolution mode at 1.43 (3D) or 1.52 (2D) μsec pixel dwell time and 4× line averaging. Airyscan alignment of the system was regularly checked during the acquisition process and raw stacks of the full ~100 μm range were finally processed by the Zeiss Airyscan processing in 3D standard mode. Additionally, single slice measurements were acquired at the indicated Z-depth in the middle of the glass tubes and processed in 2D standard airyscan mode.

## AUTHOR INFORMATION

**Corresponding Author**

Matthias Karg

E-mail: karg@hhu.de

## NOTES

The authors declare no conflict of interest.

## ACKNOWLEDGEMENTS

The authors acknowledge the DFG and the state of NRW for funding the cryo-TEM (INST 208/749-1 FUGG). M.K. acknowledges the German Research Foundation (DFG) for funding under grant KA3880/6-1. We also would like to acknowledge the Center for Advanced Imaging (CAi) at Heinrich-Heine-University Düsseldorf (HHU Düsseldorf) for providing access to the Zeiss LSM880 Airyscan system (Ref.No. DFG- INST 208/746-1 FUGG). The authors would like to thank Marco Hildebrandt (HHU Düsseldorf) for the transferred know-how for hollow glass tube handling, Marius Otten (HHU Düsseldorf) for the TEM imaging, and Dr. Virginia Carrasco Fadanelli (HHU Düsseldorf) for the kind introduction to MATLAB.

## SUPPORTING INFORMATION AVAILABLE

The following supporting information can be also downloaded at: www.mdpi.com/2310-2861/8/8/516/s1, additional synthetic details, methods and instrumentations, more detailed description of monolayer preparation, schematics and other details on SALS setup, TEM images with a higher resolution, a brief discussion on the effect of SDS, KPS, temperature and stirring on CS microgels, fitting area fraction for monolayers (PDF), two videos of acquired z-stacks on 3D samples (AVI), and a 3D model of the multi-reactor holder for small scale synthesis (STL).

# Supplementary Information

# Micron-sized silica-PNIPAM core-shell microgels with tunable shell-to-core ratio

*Keumkyung Kuk [a], Lukas Gregel [a], Vahan Abgarjan [a], Caspar Croonenbrock [a], Sebastian Hänsch [b], and Matthias Karg [a]\**

[a]Institut für Physikalische Chemie I: Kolloide und Nanooptik, Heinrich-Heine-Universität Düsseldorf, Universitätsstr. 1, 40225 Düsseldorf, Germany

[b]Center for Advanced Imaging, Heinrich Heine University Düsseldorf, Universitätsstraße 1, 40225 Düsseldorf, Germany.

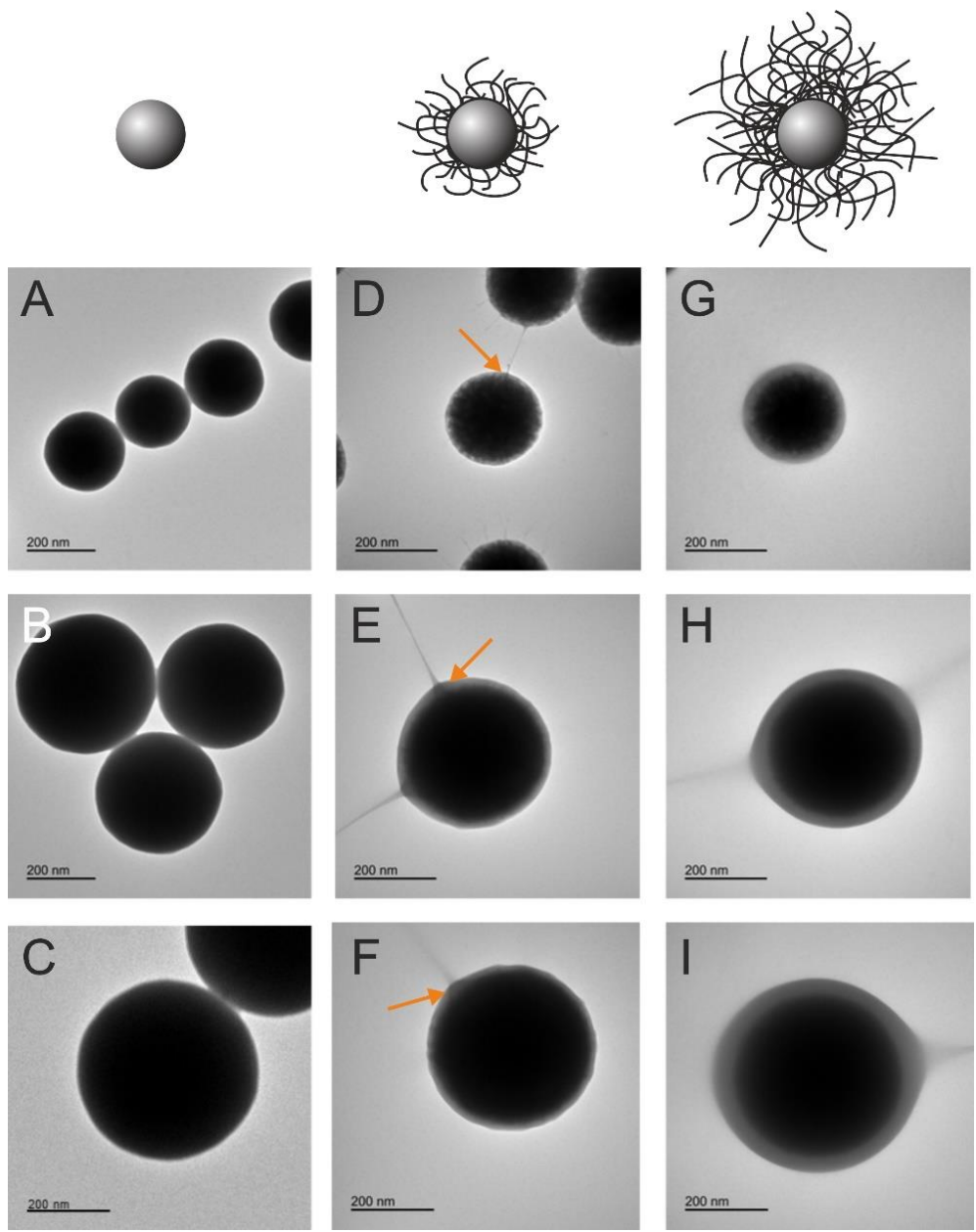

**Figure S1**. Higher magnification TEM images of CS microgels with variously sized cores: $C_{245}$ (A), $C_{388}$ (B), and $C_{455}$ (C), with thin shells: $C_{245}S_{1.68}$ (D), $C_{388}S_{2.12}$ (E), $C_{455}S_{2.11}$ (F) and with thicker shells: $C_{245}S_{2.89}$ (G), $C_{388}S_{2.62}$ (G), $C_{455}S_{2.34}$ (G).

**Influencing parameters in CS microgel synthesis at low TSC.** A surfactant, such as SDS, can be used in the synthesis to prevent the aggregation of the cores but it also influences the final size of the CS microgels. For classical microgels, the effect of surfactant is well known: It stabilizes the forming primary microgels via charge repulsion thereby driving the systems closer to good-solvent conditions. The smaller the primary microgels are, the higher is the final number of microgels, which consequently leads to smaller final size of microgels for a given amount of monomer in the system and higher homogeneity. [31-33] For CS microgels synthesized via seeded precipitation polymerization, however, the number of seeds is predetermined by the added number of cores and thus the same mechanism cannot hold. We found that CS microgels nonetheless follow the same trend as the classical microgels. **Figure S2A** shows swelling curves of CS microgels synthesized with various concentrations of SDS: 0 mM ($C_{433}S_{1.99}$-0, light green), 0.2 mM ($C_{433}S_{1.75}$-$SDS_{0.2}$, green), 2 mM ($C_{433}S_{1.63}$-$SDS_{2}$, dark green). Apart from the influence of SDS, we also investigated the role of the initiator (KPS) concentration and polymerization temperature, which has a similar influence on the size of the CS microgels, as shown in **Figure S2B**. Light blue represents 0.01 wt% KPS ($C_{433}S_{1.99}$-0), blue 0.03 wt% ($C_{433}S_{1.73}$-$KPS_{0.03}$), dark blue 0.05 wt% ($C_{433}S_{1.61}$-$KPS_{0.05}$), respectively.

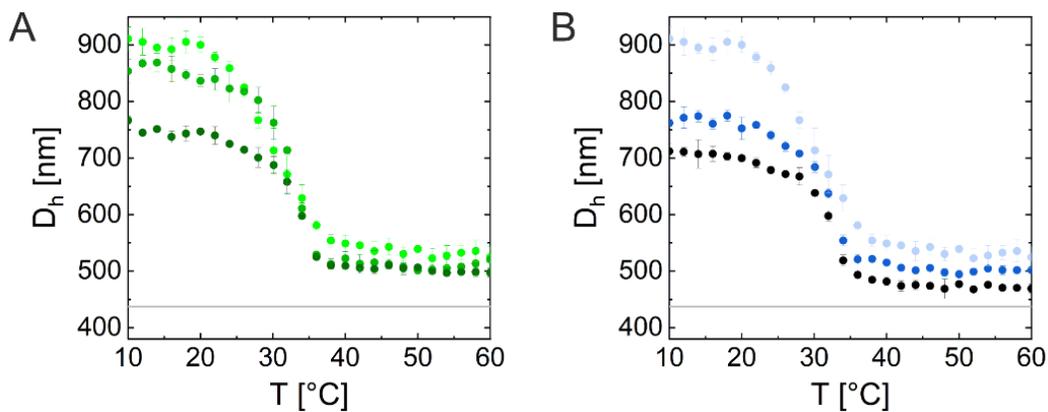

**Figure S2**. A) Swelling curves of CS microgels synthesized with increasing SDS concentrations (light green: 0 mM, green: 0.2 mM, dark green: 2 mM, grey line: core diameter by TEM. B) with increasing KPS concentrations (light blue: 0.01 wt%, blue: 0.03 wt%, dark blue: 0.05 wt%, grey line: core diameter by TEM.

The synthesis temperature is proven to exert even stronger influence on size of the microgels. For instance, over 2.5 μm sized classical microgels can be synthesized via precipitation polymerization with temperature ramp, usually at a lower temperature range. [17, 34] In contrast to the effect of surfactant or initiator, the lower temperature reduces the number of primary microgels and therefore leads to the larger microgels. **Figure S3A** shows the influence of the synthesis temperature on the size of the CS microgels. The empty pink stars, filled light red circles, red squares, and dark red triangles represent the CS microgels synthesized with a temperature ramp (45 - 65 °C, $C_{340}S_{3.36}$-$T_{45-65}$) and at fixed temperatures of 60 °C ($C_{340}S_{3.00}$-$T_{60}$), 70 °C ($C_{340}S_{2.86}$-$T_{70}$), and 80 °C ($C_{340}S_{2.72}$-$T_{80}$), respectively. As for the classical microgels, the size of CS microgels also increased with decreasing synthesis temperature. The CS microgel synthesized via temperature ramping, meanwhile, did not show a strong increase in size and exhibited considerably higher polydispersity when assembled in 2D. Additionally, more efficient stirring also seemed to improve the efficiency in PNIPAM shell growth. **Figure S3B** shows PNIPAM shell encapsulation at low TSC (0.005 g/ml) at three different synthesis temperatures: 60 °C, 70 °C, 80 °C. The filled symbols represent the syntheses stirred with a KPG stirrer with a moon-shaped stirrer blade, and the empty symbols with the egg-shaped magnetic stirring bar. The synthesis at 60 °C stirred with egg-shaped magnetic stirring bar stopped stirring an hour after the initiation due to macroscopic aggregates around the stirring bar.

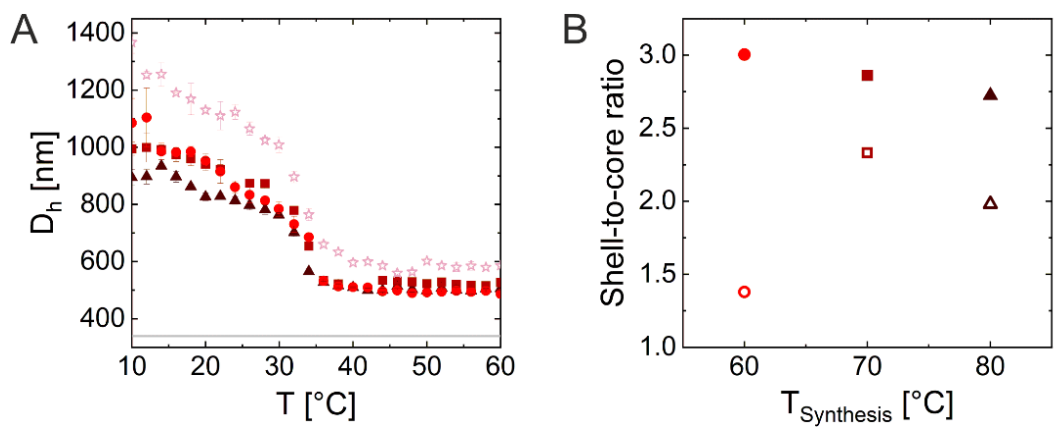

**Figure S3.** A) Swelling curves of CS microgels synthesized with decreasing temperature (empty light pink stars: 45 - 65°C filled light red circles: 60°C, red squares: 70°C, dark red triangles: 80°C, grey line: core diameter by TEM. B) Comparison between stirring by egg-shaped magnetic stirring bar (empty scatters) and KPG stirrer with moon-shaped stirrer blades (filled scatters) at various temperature: 60 °C (circles), 70 °C (squares), 80 °C (triangles).

**Monolayer preparation.** Monolayers of different CS microgels were prepared using interface-mediated self-assembly using a crystallizing dish filled with water. The air/water interface was cleaned by using an aspirator with a tip, as illustrated in **Figure S4A**. Surface pressures were measured using a Wilhelmy film balance. The CS microgels were deposited directly at the air/water interface by injection from ethanolic dispersion using a micropipette. Three different surface pressures (10, 20 and 30 mN/m) were targeted to vary the interparticle distance of the transferred monolayer. The injection was done slowly at a shallow angle while the tip was gently touching the interface until the target surface pressure was reached (**Figure S4B)**. The prepared glass slide was pushed into the water bulk phase through the monolayer close to the edge of the crystallizing dish and moved to the center. The monolayer was taken from the center at a steep angle and immediately heat-treated by a heat gun from the bottom side while horizontally held until completely dried (**Figure S4C**). The bottom side was then carefully wiped with ethanol-soaked tissue.

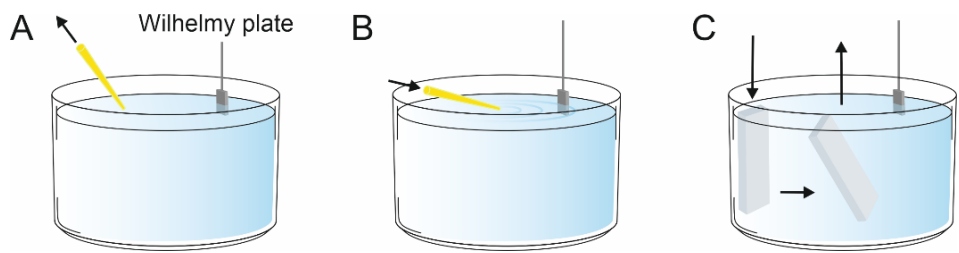

**Figure S4**. A) Air/water interface cleaning by an aspirator. B) Monolayer deposition with a micropipette at the cleaned air/water interface. C) Monolayer transfer on to the RCA cleaned glass.

**Estimation of Area fraction for 2D assembly.** Approximately 60,000 particles were counted per sample for the image analysis. The positions of the microgels and number of particles in the probed area (microscopic images) were found using ImageJ and the $D_{c-c}$ was obtained from the Gaussian fitting of the first peak of the radial distribution function. **Figure S5** shows the linear relation between $D_{c-c}^2$ and the area per particle (A/P). Note that the estimated $A_f$ is likely to be underestimated due to the fact that the microgels at the edge of the probed area are often not considered in the image analysis.

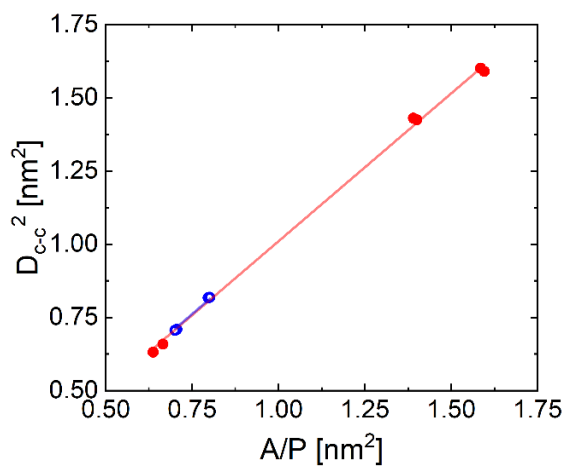

**Figure S5**. The linear fit for the calculation of $A_f$ from measured data: $C_{245}S_{2.89}$ (empty blue) and $C_{455}S_{2.34}$ (filled red).

**Table S1.** Chemicals used for the synthesis of silica cores presented in this paper.

|  | Temp. [°C] | Initial | | | Added | | $D_{TEM}$ [nm] | Dye |
|---|---|---|---|---|---|---|---|---|
|  |  | Ethanol [ml] | Water [ml] | Ammonia aq. [ml] | TEOS [ml] | Ethanol [ml] |  |  |
| $C_{245}$ | 40 | 100 | 0 | 10 | 5 | 0 | 245±18 | - |
| $C_{388}$ | 50 | 57 | 0 | 19 | 15 | 30 | 388±35 | - |
| $C_{455}$ | 60 | 90 | 0 | 20 | 30 | 10 | 455±27 | - |
| $C_{340}$ | 60 | 80 | 5 | 25 | 10 | 30 | 340±20 | Rhodamine B |
| $C_{433}$ | 50 | 56 | 16 | 28 | 10 | 40 | 433±15 | Rhodamine B |

The silica particles were synthesized via Stöber procedure. All particles were MPS modified according to the previously published protocol [24] and $C_{340}$ and $C_{433}$ were dyed with Rhodamine B.

**Table S2.** Chemicals and synthesis parameters used for the preparation of CS microgels presented in this work.

| Name | Core | | | | Core - shell | | | | | | |
|---|---|---|---|---|---|---|---|---|---|---|---|
|  | $D_{TEM}$ [nm] | TSC [g/ml] | added [ml] | $H_2O$ [ml] | NIPAM [g] | BIS [g] | KPS [g] | SDS [g] | Temp. [°C] | δ 20°C | δ 50°C |
| *$C_{433}S_{1.99}$-0 | 437 | 0.53 | 0.02 | 6 | 0.020 | 0.001 | 0.001 | 0.0000 | 80 | 1.99 | 0.00 |
| *$C_{433}S_{1.75}$-$SDS_{0.2}$ | 437 | 0.53 | 0.02 | 6 | 0.020 | 0.001 | 0.001 | 0.0003 | 80 | 1.75 | 1.10 |
| *$C_{433}S_{1.63}$-$SDS_2$ | 437 | 0.53 | 0.02 | 6 | 0.020 | 0.001 | 0.001 | 0.0035 | 80 | 1.63 | 1.11 |
| *$C_{433}S_{1.73}$-$KPS_{0.03}$ | 437 | 0.53 | 0.02 | 6 | 0.020 | 0.001 | 0.002 | 0.0000 | 80 | 1.99 | 1.13 |
| *$C_{433}S_{1.61}$-$KPS_{0.05}$ | 437 | 0.53 | 0.02 | 6 | 0.020 | 0.001 | 0.003 | 0.0000 | 80 | 1.83 | 1.18 |
| $C_{340}S_{2.72}$-$T_{80}$ | 340 | 0.11 | 2.00 | 255 | 1.002 | 0.072 | 0.026 | 0.0000 | 80 | 2.72 | 1.50 |
| $C_{340}S_{2.86}$-$T_{70}$ | 340 | 0.11 | 2.00 | 255 | 1.002 | 0.072 | 0.026 | 0.0000 | 70 | 2.86 | 1.54 |
| $C_{340}S_{3.00}$-$T_{60}$ | 340 | 0.11 | 2.00 | 255 | 1.000 | 0.072 | 0.026 | 0.0000 | 60 | 3.00 | 1.52 |
| $C_{340}S_{3.36}$-$T_{45-65}$ | 340 | 0.11 | 1.00 | 255 | 1.000 | 0.071 | 0.026 | 0.0000 | 45-65 | 3.36 | 1.85 |
| $C_{245}S_{1.26}$-C1 | 245 | 0.10 | 2.00 | 55 | 0.101 | 0.008 | 0.006 | 0.0319 | 80 | 1.26 | 1.17 |
| $C_{245}S_{1.35}$-C2 | 245 | 0.10 | 2.00 | 55 | 0.202 | 0.015 | 0.006 | 0.0314 | 80 | 1.35 | 1.19 |
| $C_{245}S_{1.44}$-C3 | 245 | 0.10 | 2.00 | 55 | 0.302 | 0.022 | 0.006 | 0.0316 | 80 | 1.44 | 1.21 |
| $C_{245}S_{1.43}$-C4 | 245 | 0.10 | 2.00 | 55 | 0.401 | 0.029 | 0.006 | 0.0314 | 80 | 1.43 | 1.14 |
| $C_{245}S_{1.45}$-C5 | 245 | 0.10 | 2.00 | 55 | 0.500 | 0.036 | 0.006 | 0.0314 | 80 | 1.45 | 1.15 |
| $C_{245}S_{1.52}$-C6 | 245 | 0.10 | 2.00 | 55 | 0.600 | 0.043 | 0.006 | 0.0315 | 80 | 1.52 | 1.19 |
| $C_{245}S_{1.57}$-C7 | 245 | 0.10 | 2.00 | 55 | 1.000 | 0.072 | 0.006 | 0.0320 | 80 | 1.57 | 1.23 |
| $C_{245}S_{1.85}$-C8 | 245 | 0.10 | 2.00 | 55 | 2.001 | 0.144 | 0.006 | 0.0322 | 80 | 1.85 | 1.39 |
| $C_{245}S_{1.68}$ | 245 | 0.11 | 0.33 | 125 | 0.076 | 0.006 | 0.013 | 0.0000 | 70 | 1.68 | 1.23 |
| $C_{245}S_{1.71}$ | 245 | 0.11 | 0.33 | 125 | 0.088 | 0.006 | 0.013 | 0.0000 | 70 | 1.71 | 1.25 |
| $C_{245}S_{2.07}$ | 245 | 0.11 | 0.33 | 125 | 0.095 | 0.006 | 0.013 | 0.0000 | 70 | 2.07 | 1.32 |
| $C_{245}S_{2.21}$ | 245 | 0.11 | 0.33 | 125 | 0.100 | 0.007 | 0.013 | 0.0000 | 70 | 2.21 | 1.30 |
| $C_{245}S_{2.49}$ | 245 | 0.11 | 0.33 | 125 | 0.106 | 0.008 | 0.013 | 0.0000 | 70 | 2.49 | 1.37 |
| $C_{245}S_{2.89}$ | 245 | 0.11 | 0.33 | 125 | 0.265 | 0.018 | 0.013 | 0.0000 | 70 | 2.89 | 1.74 |
| $C_{245}S_{3.07}$ | 245 | 0.11 | 0.33 | 125 | 0.404 | 0.027 | 0.013 | 0.0000 | 70 | 3.07 | 1.82 |
| $C_{388}S_{1.75}$ | 388 | 0.37 | 1.00 | 125 | 0.192 | 0.014 | 0.013 | 0.0000 | 70 | 1.75 | 1.26 |
| $C_{388}S_{2.12}$ | 388 | 0.37 | 1.00 | 125 | 0.304 | 0.021 | 0.013 | 0.0000 | 70 | 2.12 | 1.38 |
| $C_{388}S_{2.43}$ | 388 | 0.37 | 1.00 | 125 | 0.383 | 0.027 | 0.013 | 0.0000 | 70 | 2.43 | 1.42 |
| $C_{388}S_{2.62}$ | 388 | 0.37 | 1.00 | 125 | 0.502 | 0.035 | 0.013 | 0.0000 | 70 | 2.62 | 1.51 |
| $C_{455}S_{1.25}$ | 455 | 0.20 | 0.36 | 125 | 0.086 | 0.006 | 0.013 | 0.0000 | 70 | 1.25 | 1.11 |
| $C_{455}S_{2.11}$ | 455 | 0.20 | 0.36 | 125 | 0.229 | 0.016 | 0.013 | 0.0000 | 70 | 2.11 | 1.32 |
| $C_{455}S_{2.23}$ | 455 | 0.20 | 0.36 | 125 | 0.502 | 0.037 | 0.013 | 0.0000 | 70 | 2.23 | 1.46 |
| $C_{455}S_{2.30}$ | 455 | 0.20 | 0.36 | 125 | 0.601 | 0.041 | 0.013 | 0.0000 | 70 | 2.30 | 1.45 |
| $C_{455}S_{2.25}$ | 455 | 0.20 | 0.36 | 125 | 0.751 | 0.052 | 0.013 | 0.0000 | 70 | 2.25 | 1.51 |
| $C_{455}S_{2.34}$ | 455 | 0.20 | 0.36 | 125 | 1.003 | 0.070 | 0.013 | 0.0000 | 70 | 2.34 | 1.41 |

*****Small scale synthesis**: See the next paragraph below along with Figure S6 for more details.

**Small scale synthesis.** 3 - 6 batches were synthesized at a time in 12 ml round bottom glass centrifuge tubes. The tubes were held by a 3D printed holder at around 45° angle, as depicted in **Figure S6**. The reaction mixtures were prepared in stock, purged with argon for 1 hour and heated up to 80°C while stirring with a winged magnetic stirrer bar. The polymerization was allowed to proceed for at least 3 hours with a constant flow of argon. The dispersions were purified by repeated centrifugation and re-dispersion in water without filtration.

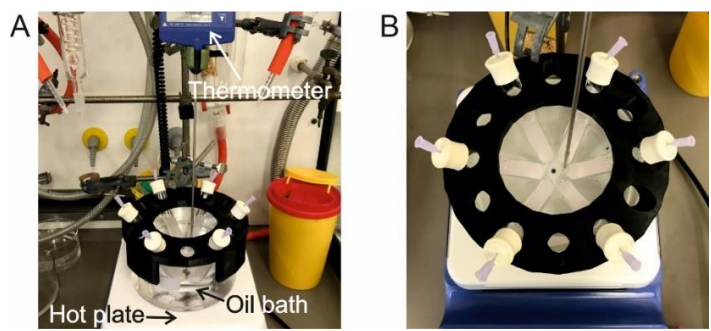

**Figure S6.** Small scale synthesis setup. A) Side view of the setup. B) Top view.

**Small-angle light scattering (SALS).** Laser diffraction patterns were recorded by a self-built setup, see **Figure S7**. A blue laser (MediaLas, LDM-20-405, 20 mW, 405 nm) was used as a light source. The diameter of the laser was reduced to 1.19 mm by a beam expander, which consists of two lenses, a pinhole and an iris (Thorlabs). The sample was placed on 3D printed holders. A CCD camera (Thorlabs, DCU223C-MVL6WA) and a paper screen as a detector, the images were captured in dark. The primary beam blocked by a 3D printed beam stop during measurement. The pixel-to-mm ratio was calibrated by using millimetre paper on the paper screen.

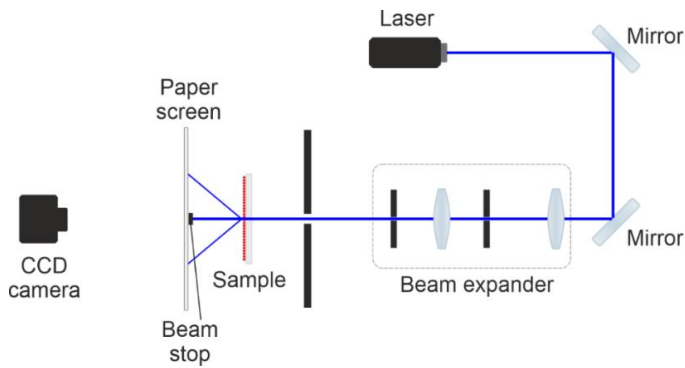

**Figure S7**. Schematic depiction of the SALS setup that was used to record diffraction patterns from monolayer samples.